\documentclass[aps,prc,twocolumn,floatfix,showpacs,amsmath,amssymb]{revtex4}
\usepackage{graphicx}
\usepackage{dcolumn}
\usepackage{color} 
  \def\nuc#1#2{\relax\ifmmode{}^{#1}{\protect\text{#2}}\else${}^{#1}$#2\fi}
  \def\itnuc#1#2{\setbox\@tempboxa=\hbox{\scriptsize\it #1}
    \def\@tempa{{}^{\box\@tempboxa}\!\protect\text{\it #2}}\relax
    \ifmmode \@tempa \else $\@tempa$\fi}
  
  \newcommand{\beq}{\begin{equation}}
  \newcommand{\eeq}{\end{equation}}
  \newcommand{\bea}{\begin{eqnarray}}
  \newcommand{\eea}{\end{eqnarray}}

  \newcommand{\fig}[1]{Fig.~\ref{#1}}
  \newcommand{\co}{(Color online)}

  \newcommand{\lisi}{\nuc{6}{Li}}
  \newcommand{\beni}{\nuc{9}{Be}}
  \newcommand{\nm}{\ensuremath{N_\mathrm{max}}}
  \newcommand{\ho}{\ensuremath{\hbar \Omega}}
  \newcommand{\nn}{\ensuremath{N\!N}}
  
 \newcommand{\cdb}{CDB2k}

\begin{document}

\title{Converging sequences in the \emph{ab initio} no-core shell model}
\author{C. Forss\'en}
\email[]{c.forssen@fy.chalmers.se}
\affiliation{Department of Fundamental Physics, Chalmers University of Technology,
  412 96 G\"oteborg, Sweden}
\author{J. P. Vary}
\affiliation{Department of Physics and Astronomy, Iowa State University, 
Ames, IA  50011, USA}
\author{E. Caurier}
\affiliation{Institut de Recherches Subatomiques
            (IN2P3-CNRS-Universit\'e Louis Pasteur)\\
            Batiment 27/1,
            67037 Strasbourg Cedex 2, France}
\author{P. Navr\'atil}
\affiliation{Lawrence Livermore National Laboratory, P.O. Box 808, L-414, 
Livermore, CA  94551, USA}
\date{\today}
\begin{abstract}
We demonstrate the existence of multiple converging sequences in the
\emph{ab initio} no-core shell model. By examining the underlying
theory of effective operators, we expose the physical foundations for
the alternative pathways to convergence.  This leads us to propose a
revised strategy for evaluating effective interactions for $A$-body
calculations in restricted model spaces. We suggest that this strategy
is particularly useful for applications to nuclear processes in which
states of both parities are used simultaneously, such as for transition
rates.  We demonstrate the utility of our strategy with large-scale
calculations in light nuclei.
\end{abstract}
\maketitle
%
\section{\label{sec:intro}Introduction}
The \emph{ab initio} no-core shell model (NCSM) is a method to solve the
full $A$-body problem for a system of non-relativistic particles that
interact by realistic two- plus three-body forces. A particular feature
of the method is the use of effective interactions appropriate for the
large, but finite, harmonic-oscillator (HO) model spaces employed in the
calculations. Over the last few years, the NCSM method has been
established as a very valuable tool in nuclear physics, aiming for an
exact description of nuclear structure starting from the fundamental
interaction between nucleons~\cite{nav00:84,nav00:62,nav01:87}.

The objective of this paper is to highlight the existence of multiple
converging sequences in the NCSM, and the opportunities that this
property provides in practical applications.
In particular, we will discuss and illustrate a specific converging
sequence that relies on the implicit freedom in the choice of
model-space cutoff when constructing the cluster-approximated effective
interactions used in the NCSM.
These effective interactions are intended to take into account effects
of configurations outside the model space. In general, the effective
interaction is derived from the underlying realistic inter-nucleon
potential by a unitary
transformation~\cite{pro64:30,suz80:64,suz82:68,suz83:70,suz94:92}. The
procedure aims to reproduce exactly a subset of the eigenvalues to the
original Hamiltonian in the finite model space. The effective
interaction, in principle, becomes an $A$-body operator but is, in
practice, usually approximated at the $a$-body operator level, where $a
< A$. This cluster approximation generates, e.g., a dependence on the
choice of HO frequency, \ho. However, the construction guarantees
convergence to the exact solution, for any value of \ho, as the size of
the model space increases.  This guarantee holds independent of the
relationship between the model space used to evaluate the cluster
approximation and the model space used to carry out the diagonalization
of the resulting $A$-body operator, as long as both increase.  In
addition, we note that for any finite model space of dimensionality
$d_P$, and in the limit $a \to A$, we obtain exact solutions for $d_P$
states of the full problem, with flexibility for the choice of physical
states subject to certain conditions~\cite{via01:42}. Thus, there is an
arbitrary number of sequences in the NCSM that all converge, in
principle, to the same result.

The implicit freedom in generating approximated effective interactions,
as mentioned above, will be introduced in section~\ref{sec:theor} in
connection to a brief description of the NCSM formalism. The strategy
that we propose to employ for certain applications has several
advantageous properties compared to the traditional strategy for
evaluating effective interactions. These properties will be discussed in
section~\ref{sec:conve}, while the features and applicability of the
revised strategy will be illustrated in section~\ref{sec:resul}. The
ideas presented in this paper are substantiated by large-scale
calculations for light nuclear systems. However, we note that our
revised strategy was already utilized in two recent papers applying the
NCSM to heavier systems (with $A = 47 - 49$) in restricted many-body
model spaces~\cite{var06:0607041,var06:0612022}. Conclusions and
perspectives are presented in section~\ref{sec:concl}.
\section{\label{sec:theor}Theory}
The goal is to solve the $A$-body Schr\"odinger equation with an
intrinsic Hamiltonian $H_A= T_\mathrm{rel} + {\cal V}$, where
$T_\mathrm{rel}$ is the relative kinetic energy and ${\cal V}$ is the
sum of two-body nuclear and Coulomb interactions. The NCSM method also
allows for the inclusion of three-body forces~\cite{mar02:66}.  However,
while realistic three-nucleon forces have been shown to be important in
obtaining the nuclear spectra~\cite{nav03:68,mar02:66,pie02:66} and in
describing electromagnetic and weak form factors~\cite{hay03:91}, we
consider only two-body interactions in this study.  We solve the
many-body problem in a finite HO basis space and the usual approach is
therefore to derive a model-space dependent effective Hamiltonian. For
this purpose, we perform a unitary transformation of the Hamiltonian, in
the spirit of Da Providencia and Shakin~\cite{pro64:30} and Lee, Suzuki
and Okamoto~\cite{suz80:64,suz82:68,suz83:70,suz94:92}, which is able to
accommodate short-range correlations. However, the very first step is to
add a center-of-mass (CM) HO Hamiltonian to the intrinsic
Hamiltonian. In the full Hilbert space the added $H_\mathrm{CM}^\Omega$
term has no influence on the intrinsic properties. When we introduce our
cluster approximation below, the added $H_\mathrm{CM}^\Omega$ term
generates a real dependence on the choice of HO frequency, but it also
facilitates faster convergence to exact results with increasing basis
size.

Note that even if the original Hamiltonian contained just one- and
two-body terms, the transformed Hamiltonian $\mathcal{H}$ contain up to
$A$-body terms. Obtaining the exact transformation operator is
equivalent to solving the initial problem, which would make the
procedure impractical. Therefore, we introduce the cluster
approximation. The approximation consists in obtaining the effective
interaction from the decoupling condition between the model space ($P$)
and the excluded space ($Q$) for the $a$-body problem, where $a \le A$,
and then using the effective interaction thus obtained in the desired
$A$-body problem. See, e.g., Ref.~\cite{nav00:84,nav00:62,nav00:61} for
details on the procedure. This approximation introduces a real
dependence on the oscillator parameter \ho. The resulting effective
$a$-body effective interactions also depends on the nucleon number $A$
and on \nm, the maximum $A$-body HO excitation energy defining the model
space.  The usual approach to handle the \ho-dependence is to search for
a range of \ho\ values over which the results are weakly
\ho-dependent. This empirical choice then corresponds to the optimal HO
frequency for the system.

There are two limiting cases of the cluster approximation: First, when
$a \to A$, the solution becomes exact; a higher-order cluster is a
better approximation and was shown to increase the rate of
convergence~\cite{nav03:68}. Second, when $P \to 1$, the effective
interaction approaches the bare interaction; and as a result, the
effects of the cluster approximation can be minimized by increasing as
much as possible the size of the model space.

In this work, we present results obtained at the two-body cluster
level. In practice, the exact (to numerical precision) solutions for the
$a=2$ cluster are obtained in basis spaces of several hundred \ho\ in
each relative motion \nn\ channel. The resulting effective Hamiltonian,
now consisting of a relative two-body operator, and the subtracted pure
$H_\mathrm{CM}^\Omega$ term introduced earlier, is then inserted into an
m-scheme Lanczos diagonalization process to obtain the $A$-body, P-space
eigenvalues and eigenvectors.  The evaluation of the $A$-nucleon
Hamiltonian and its diagonalization is a highly nontrivial problem due
to the very large dimensions we encounter. For the present work, we
performed the many-body calculation with two completely independent
shell-model codes: a specialized version of the code
\textsc{Antoine}~\cite{cau99:30}; and the many-fermion dynamics
\textsc{MFD} shell-model code~\cite{var92:unp}. At the diagonalization
stage we also add the Lawson projection term, $\beta
(H_\mathrm{CM}^\Omega - \frac{3}{2}\ho)$ (with $\beta$ being a large
positive coefficient) to separate the physically interesting states with
0s CM motion from those with excited CM motion. We retain only the
eigenstates with pure 0s CM motion when evaluating observables. All
observables that are expressible as functions of relative coordinates,
such as the rms radius and radial densities, are then evaluated free of
CM motion effects.
\section{\label{sec:conve}Converging Sequences}
We define our P-space to consist of all $A$-body configurations in the
oscillator basis with total oscillator energy less than or equal some
cutoff value $(N_m +3 A / 2)\ho$, where $N_m = N_\mathrm{min} + \nm$,
and $N_m$ is the sum of $2n+l$ values of the occupied single-particle
states in the configuration. $N_\mathrm{min}$ is the minimum value
required by the Pauli principle. As an example, $N_\mathrm{min} = 2$ for
\lisi. The P-space is equally described by the cutoff parameter \nm,
that begins with 0. We note that the usual convention is to solve only
for states whose parity corresponds to the configurations in the maximum
subspace governed by \nm.  The opposite-parity states are then obtained
in the $\nm + 1$ model space. The corresponding two-body cluster model
space, $P_2$, is defined by the range of two-body states encountered in
the P-space. This implies that the relative $(n,l)$ states are
restricted by the condition $2n+l \leq N_m - N_\mathrm{spsmin}$, where
$N_\mathrm{spsmin}$ denotes the minimum possible number of HO
excitations of the $(A-2)$ spectators. Consider, e.g., a \lisi\
calculation in the $\nm = 4$ model space. In this case $N_m = 6$ and
$N_\mathrm{spsmin} = 0$ (since the 4 spectator nucleons can all fit in
the 0s shell), which leads to a two-body cutoff at $2n+l \leq 6$.

Due to our cluster approximation a dependence of our results on \nm\ and
on \ho\ arises. For a fixed cluster size $a$, the smaller the basis space,
the larger the dependence on \ho. The residual \nm\ and \ho\ dependences
can be used to infer the uncertainty in our results arising from the
neglect of effective many-body interactions.  

The usual NCSM strategy has been to evaluate $H_\mathrm{eff}$ for each
$A$-body model space. This strategy leads to the use of separate
$H_\mathrm{eff}$ for positive- and negative-parity states. The
convergence of, e.g., the energy spectrum of natural-parity states is
then observed by performing calculations in a sequence of model spaces
$\nm = 0, 2, 4, \ldots$, with corresponding effective interactions,
while unnatural-parity states are obtained in the $\nm = 1, 3, 5,
\ldots$ sequence. However, the theory guarantees convergence to the
exact results of any sequence that follows a simple rule and brings us
to the full $A$-body Hilbert space ($P \to 1$) as the model space is
increased. In particular, we can consider the following rule for
evaluating the effective Hamiltonian:
\beq
  N_\mathrm{max,eff} = \nm + N_\mathrm{shift}, 
\eeq
where convergence is obtained for $\nm \to \infty$, and where
$N_\mathrm{shift} = 0$ is the conventional choice. We will instead
consider the combination of choosing $N_\mathrm{shift} = 1$ for
natural-parity states and $N_\mathrm{shift} = 0$ for unnatural-parity
states. This choice implies that we will have the same $H_\mathrm{eff}$
for both positive- and negative-parity states in adjoining model spaces,
e.g., the $N_\mathrm{max,eff} = 1$ effective Hamiltonian will be used in
both the $\nm = 0$ and $\nm = 1$ model spaces. The logic for the revised
strategy stems from four considerations: (1) either strategy will
converge to the exact result in sufficiently large model spaces; (2) the
relative position of positive- and negative-parity states will converge
faster when the same effective Hamiltonian is used for both of them; (3)
for adjoining spaces in heavier systems, the predominant sets of
pairwise interactions are in the same configurations with just one pair
at a time shifting to the larger space; and (4) for electromagnetic
transitions between states of opposite parity, the theory of the
corresponding effective operators will be simplified. In addition to
these four considerations, the revised strategy simplifies our work to
compute $H_\mathrm{eff}$ since it is required only for every other
increment in the basis space, such as $N_\mathrm{max,eff} = 1, 3, 5,
\ldots$, to evaluate the converging sequence.

In the next section we will explicitly illustrate the first two points
by presenting results from large-scale calculations for two p-shell
nuclear systems: \lisi\ and \beni. The third point was highlighted in
two recent papers applying the NCSM to systems with $A = 47 - 49$ in
restricted many-body model spaces~\cite{var06:0607041,var06:0612022}. In
those papers, the authors particularly argued that the bulk of the
binding should not be altered in proceeding from a 0\ho\ to a 1\ho\
model space in $A=48$, suggesting the same $H_\mathrm{eff}$ is
preferred. Finally, the fourth consideration is useful for applications
of the NCSM formalism involving transitions between states of different
parities; and consequently for the future description of low-energy
reactions such as electric dipole radiative capture processes.
%
\section{\label{sec:resul}Results}
\subsection{\label{sec:resulA}%
Converging sequences: different $N_\mathrm{shift}$}
We have performed calculations for \lisi\ up through the 16\ho\ $(\nm =
16)$ model space, and for \beni\ up through the 10\ho\ $(\nm = 10)$
model space. The maximum dimensions of the encountered model spaces are
$d_P = 7.9 \times 10^8$ and $d_P = 5.4 \times 10^8$ for \lisi\ and
\beni, respectively.

Most of the results presented in this paper are obtained using the
non-local CD-Bonn 2000 (\cdb) potential~\cite{mac01:63}, which is a
charge-dependent \nn\ interaction based on one-boson exchange. The
off-shell behavior of the \cdb\ interaction differs from local
potentials which leads to larger binding energies in nuclear few-body
systems. Still, the \cdb\ gives underbinding of nuclear many-body
systems as is typical for standard high-precision \nn\ interactions. In
general, the \cdb\ potential applied in the NCSM gives a good
convergence rate and a weak HO frequency dependence.

A more recent realistic, nonlocal \nn\ interaction, obtained through an
inverse scattering analysis of the \nn\ data, is able to provide
reasonable binding energies of p-shell nuclei and is called the "JISP16"
interaction~\cite{shi07:644}.  We will use this interaction in
Sec.~\ref{sec:resulB} to illustrate the value of exponential fits to
converging sequences in the NCSM.

The optimal HO frequency for each isotope was found by performing a
series of calculations for a number of different frequencies and model
spaces. We searched for the region in which the dependence on \ho\ is
minimal; and we selected this frequency (from the calculation in the
largest model space) to use in the more detailed investigation. In this
way we found that the optimal frequencies for the \cdb\ interaction are
$\ho = 10$~MeV for \lisi\ and $\ho = 12$~MeV for \beni\ (see also
Ref.~\cite{for05:71}). Since we will also use the $\nm+1$ effective
Hamiltonian for natural-parity states, diagonalized in the \nm\ model
space, we verified by explicit calculations that the same choices of
optimal frequencies apply.

Note that the choice of optimal frequency is dependent on the model
space used in the calculations. Consequently, a larger HO frequency
($\ho = 13$~MeV) was employed in an earlier NCSM study of \lisi, in
which the calculations were limited to $\nm \le 10$~\cite{nav01:87}.

\begin{figure}[htb!]
  \centering
  \includegraphics*[width=80mm,clip=true]{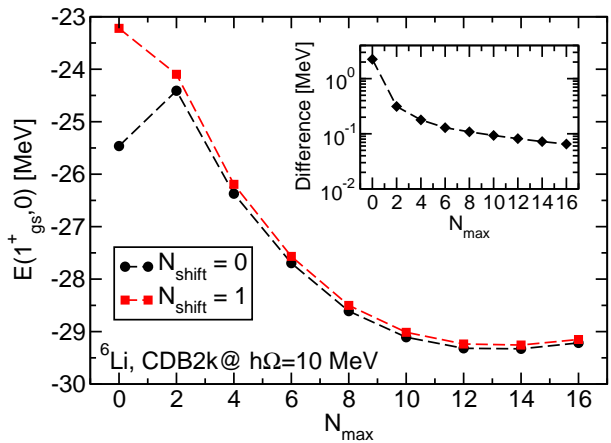}
  \caption{\co\ Ground-state binding energy of \lisi\ calculated with the
    \cdb\ Hamiltonian with $\ho=10$~MeV. Results obtained using
    two-body effective interactions generated with $N_\mathrm{shift} =
    0$ and  $N_\mathrm{shift} = 1$ are compared.%
  \label{fig:6ligs}}
\end{figure}

The ground-state binding energy of \lisi, as a function of increasing
model space \nm, is presented in \fig{fig:6ligs}. The results obtained
using the traditional cutoff for evaluating the effective interaction,
$N_\mathrm{max,eff} = \nm$, is represented by black circles, while the
results obtained with $N_\mathrm{max,eff} = \nm + 1$ is represented by
red squares.
A number of observations can be made: First, these calculations
demonstrate that the NCSM with the cluster-approximated effective
interaction is not a variational approach. Convergence is not
necessarily from above. Second, the difference between the two
strategies for computing the effective interactions is the largest for
small model spaces. The choice $N_\mathrm{max,eff} = \nm + 1$ reduces
the excursions in the renormalized results in the lowest basis spaces as
a function of $\nm$. Third, both strategies converge to the same value
as $\nm \to \infty$. The binding-energy difference between the two
calculations is plotted on a semi-logarithmic scale in the inset of
\fig{fig:6ligs}. We find that the convergence rate towards zero
difference is exponential. The observed difference in our largest model
space ($\nm = 16$) is less than 70~keV (corresponding to $\sim 0.2$\% of
the total binding energy).

\begin{figure}[hbt]
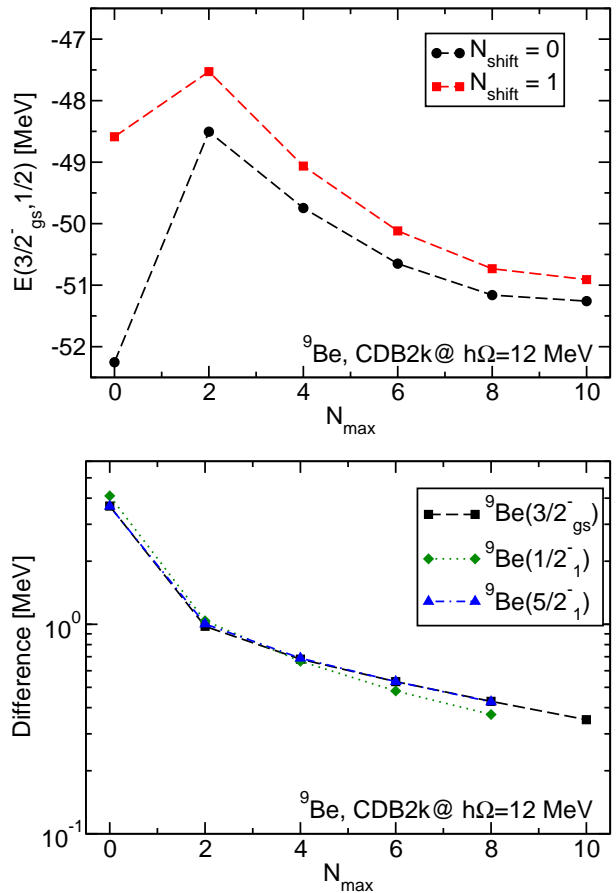

  \includegraphics*[width=80mm,clip=true]
  		   {fig2a_be9cdb_gs_Nmax1.eps}\\[2ex]
  \includegraphics*[width=80mm,clip=true]
  		   {fig2b_be9cdb_1_3_5_Nmax1.eps}\\
    \caption{\co\ Upper panel: Ground-state binding energy of \beni\
    calculated with the \cdb\ Hamiltonian with $\ho=12$~MeV. Results
    obtained using two-body effective interactions generated with
    $N_\mathrm{shift} = 0$ and $N_\mathrm{shift} = 1$ are
    compared. Lower panel: Difference between binding energies computed
    with $N_\mathrm{shift} = 0$ and $N_\mathrm{shift} = 1$ for the first
    three natural-parity states of \beni.%
    \label{fig:9begs}}
\end{figure}

We present the results of a similar comparison for \beni\ in
\fig{fig:9begs}.
The spectroscopy and various properties of this nucleus were extensively
studied within the NCSM approach in Ref.~\cite{for05:71}. The same
observations as for \lisi\ concerning the convergence rate apply for
this case, although with $A = 9$ we are slightly further from converged
results. Still, we find an exponential convergence rate towards zero
difference between the two strategies, as shown in the lower panel of
\fig{fig:9begs}. We studied three different, low-lying natural-parity
states and found the same convergence pattern. For the ground state
calculated in the $\nm = 10$ model space, the difference in computed
binding energy is 350~keV ($\le 0.7$\% of the total binding energy).

As an additional remark, we note that the eigenenergies obtained with
the $\nm + 1$ effective interaction are always above the ones obtained
with the \nm\ effective interaction. This observation persists for other
systems that we have studied. An indication of the underlying reason for
this observation can be obtained from the following argument: The $\nm +
1$ effective interaction can be diagonalized in the $\nm + 1$ model
space, but we choose to diagonalize it in the \nm\ model space. Thus, we
have a truncation of the ``available'' basis space so the result will be
above the one with the same Hamiltonian in the full ``available'' space.
This argument does not prove though, that it will always be above the
results in the \nm\ basis space with the \nm\ effective Hamiltonian.

\begin{figure*}[hbt]
    \begin{minipage}{0.96\textwidth}
      \begin{minipage}[t]{0.49\textwidth}
        \centering
        \mbox{}\\
        \includegraphics*[width=80mm,clip=true]
  		       {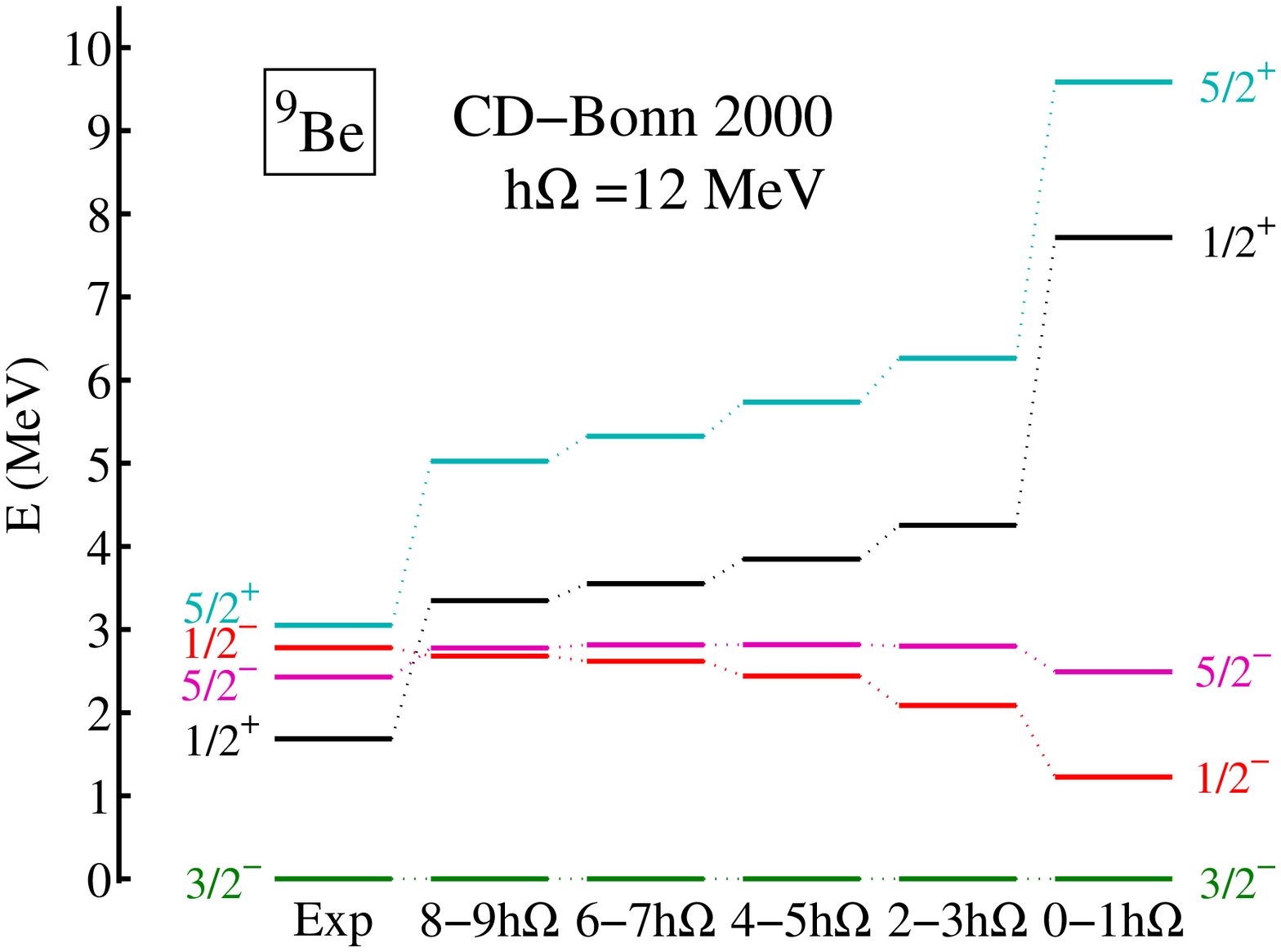}\\
        \mbox{}
      \end{minipage}
    \hfill
      \begin{minipage}[t]{0.49\textwidth}
        \centering
        \mbox{}\\
        \includegraphics*[width=80mm,clip=true]
  		       {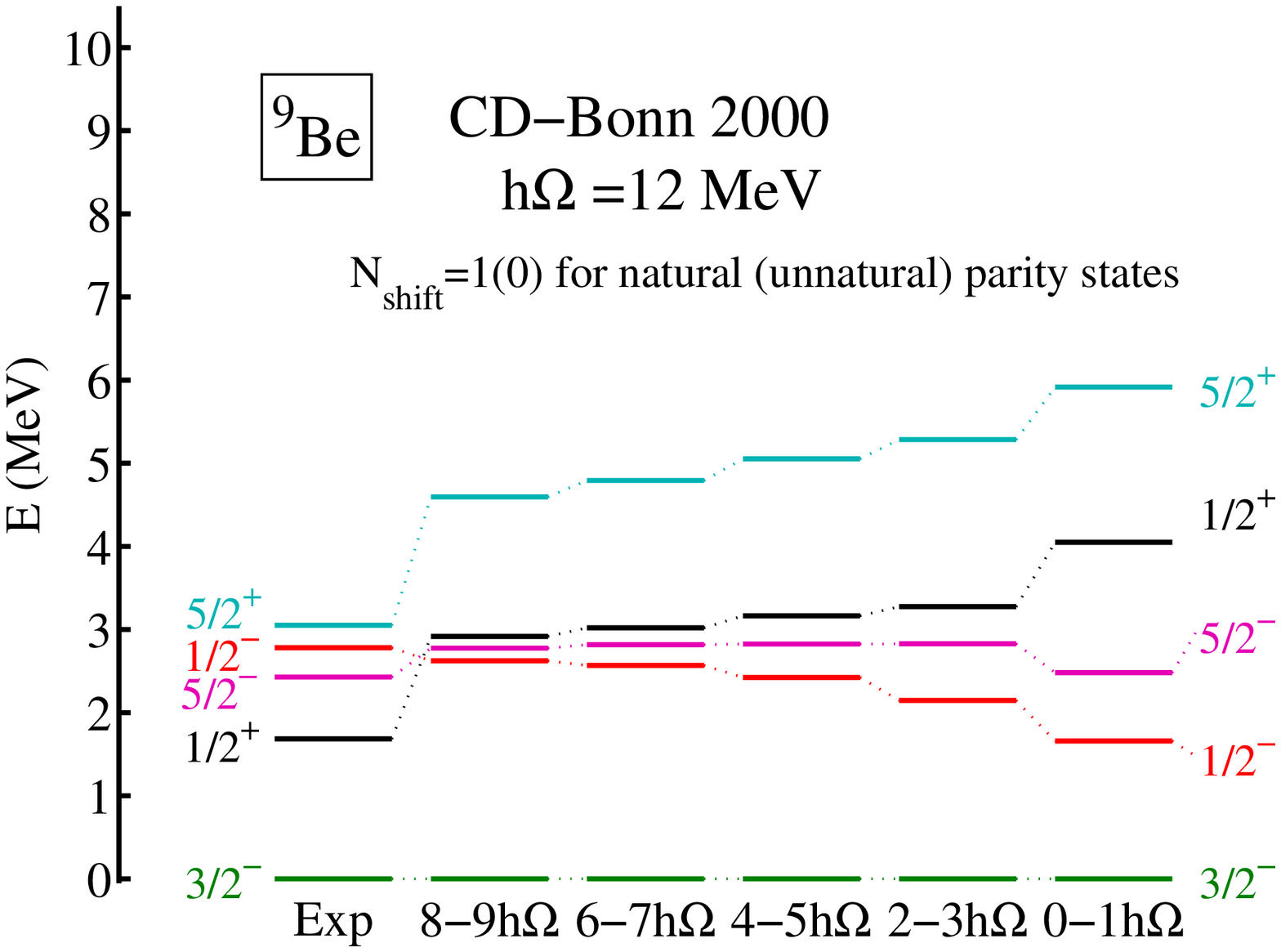}\\
        \mbox{}
      \end{minipage}
    \end{minipage}
    \caption{\co\ Spectrum of natural- and unnatural-parity states of
    \beni\ calculated with the \cdb\ Hamiltonian with $\ho=12$~MeV. The
    natural-parity states are obtained using two-body effective
    interactions generated with $N_\mathrm{shift} = 0$ (left panel) or
    $N_\mathrm{shift} = 1$ (right panel).%
    \label{fig:9bespec}}
\end{figure*}

Up until now we have demonstrated that the two strategies converge to
the same value, and that the $\nm + 1$ effective interaction gives
better renormalized results in the smallest ($0\ho$) model space. We
also want to show that the relative position of positive- and
negative-parity states converges faster when they are obtained by
diagonalizing the same effective Hamiltonian. This claim is supported by
the results presented in \fig{fig:9bespec}. In this figure, the combined
spectrum of natural- and unnatural-parity states of \beni\ is plotted.
The two panels show the evolution of the spectra with increasing model
spaces for the two strategies: (left) different effective Hamiltonians
for each separate model space; and (right) the same effective
Hamiltonian for each pair of model spaces, $\nm = (0\!-\!1), \:
(2\!-\!3), \: (4\!-\!5), \ldots$. The evaluated experimental
spectrum~\cite{til04:745} is shown in the leftmost column. Indeed, we
find that the relative position of positive- and negative-parity states
is better described already for small model spaces with the $\nm + 1$
effective interaction being used for the natural-parity states. 

\begin{figure}[htb!]
  \centering
  \includegraphics*[width=80mm,clip=true]
		  {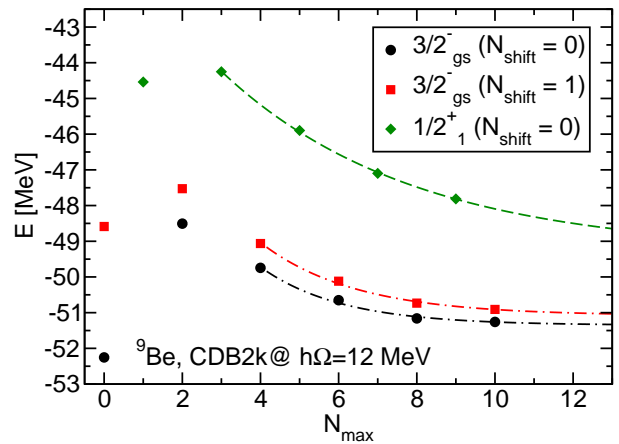}
  \caption{\co\ Basis size dependence of the calculated
    $E_(3/2_\mathrm{gs}^-)$ and $E_(1/2_1^+)$ binding energies. The
    \cdb\ Hamiltonian with $\ho=12$~MeV is used. The binding energy of
    the (natural-parity) ground state is obtained using two-body
    effective interactions generated with $N_\mathrm{shift} = 0$ (black
    circles) and $N_\mathrm{shift} = 1$ (red squares). See text for details.%
  \label{fig:9bebinding}}
\end{figure}

\begin{figure}[htb!]
  \centering
  \includegraphics*[width=80mm,clip=true]
		  {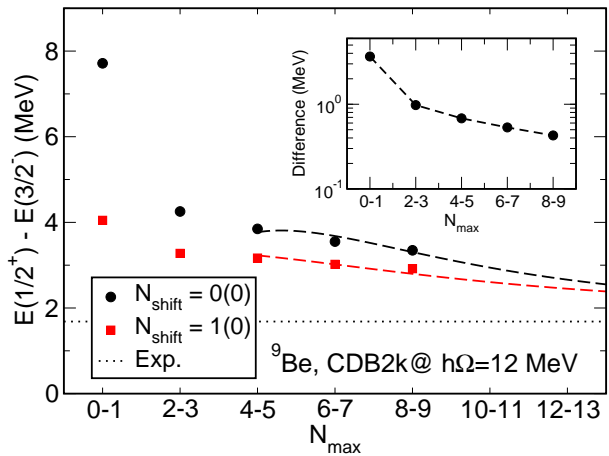}
  \caption{\co\ Basis size dependence of the calculated $E_x(1/2_1^+)$
    excitation energy relative to the lowest negative-parity state. The
    \cdb\ Hamiltonian with $\ho=12$~MeV is used. The binding energy of
    the (natural-parity) ground state is obtained using two-body
    effective interactions generated with $N_\mathrm{shift} = 0$ (black
    circles) and $N_\mathrm{shift} = 1$ (red squares).%
  \label{fig:9beexc}}
\end{figure}

Exponential fits to the calculated eigenenergies of the first natural-
and unnatural-parity states are shown in \fig{fig:9bebinding}.  The two
sequences of calculations of $E(3/2_\mathrm{gs}^-)$ (with
$N_\mathrm{shift}=0$ and $N_\mathrm{shift}=1$, respectively) constitute
an example of a series of converging sequences that converge to the same
exact result.  Therefore, we carry out a constrained fit of the two data
sets. We use exponential functions and the constraint for the fit is to
have a common asymptote to both sequences.  The fit produces a common
asymptote of -51.40 MeV with uncertainty of about 200 keV.  In carrying
out this constrained fit, the chi-square weights were based on the local
derivative with respect to $\nm$ so that results with minimal
sensitivity to $\nm$ receive greater weight. The $\nm = 0-2$ results
were excluded from the fit. We attempted alternative functional forms,
such as $\alpha + \beta \nm^{-\gamma}$, but found none to be as
successful as the constrained exponential fits shown in
Fig.~\ref{fig:9bebinding}.
Finally, we present in \fig{fig:9beexc} a plot of the excitation energy
of the first unnatural-parity state in \beni, relative the
natural-parity ground state, as a function of the model space size for
the two different strategies.  These results illustrate rather clearly
that the relative position of natural- and positive-parity states is
better described in the revised strategy where the same effective
Hamiltonian is being used for diagonalization in both \nm\ and $\nm + 1$
model spaces.
The dashed lines correspond to the differences between the eigenenergy
fits of \fig{fig:9bebinding}. The common asymptote is $E_x(1/2_1^+) =
2.10$~MeV. The difference in excitation energy between the two
calculations is plotted on a semi-logarithmic scale in the inset of
\fig{fig:9beexc}.
\subsection{\label{sec:resulB}Converging sequences: different \ho}
\begin{figure}[htb!]
  \centering
  \includegraphics*[width=80mm,clip=true]
		  {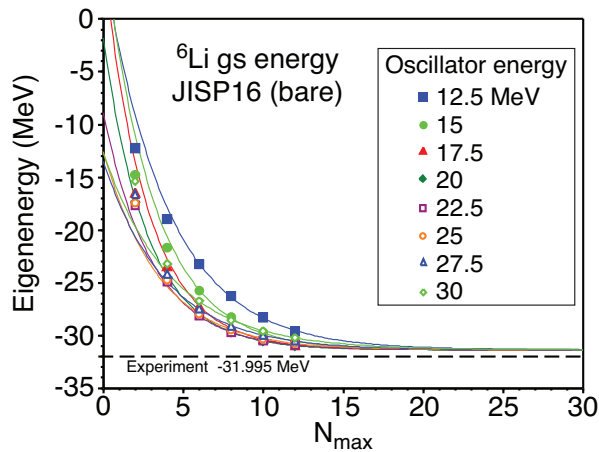}
  \caption{\co\ Ground-state binding energy of \lisi\ calculated with the
    JISP16 Hamiltonian for different HO frequencies
    ($\ho=12.5-30$~MeV). The curves are extrapolated using a constrained
    exponential fit as described in the text.%
  \label{fig:6liho}}
\end{figure}

Additional examples of series of converging sequences in the \emph{ab
initio} NCSM can be found. In the following we present a series of
calculations of the ground state eigenvalue of \lisi\ using the JISP16
realistic \nn\ interaction \cite{shi07:644}. Fig.~\ref{fig:6liho}
displays the results for the bare interaction as a function of $\nm$ up
through $\nm = 12$.  
In this case the series consists of the eight sequences of calculations
for increasing model spaces performed at different HO frequencies,
ranging from $\ho = 12.5-30$~MeV. The $\nm = 0$ results are
suppressed. These calculations should all converge to the same result as
the dependence on the choice of HO frequency disappears with increasing
model space. In addition, the procedure is variational due to the use of
bare interactions. We carry out a constrained fit using exponential
functions for each sequence at fixed \ho\ using only the $\nm = 6$
through $\nm = 12$ results. The constraint for the fit is to have a
common asymptote to all sequences.  The uniformity of the convergence is
striking and produces a common asymptote of -31.33 MeV with uncertainty
of about 100 keV. Again, the chi-square weights used in the fit were
based on the local derivative with respect to $\nm$ so that results with
minimal sensitivity to $\nm$ receive greater weight.  We attempted
alternative functional forms but found none to be as successful as the
constrained exponential fits shown in Fig.~\ref{fig:6liho}.
%
\section{\label{sec:concl}Conclusion}
The \emph{ab initio} NCSM is generally characterized by providing very
fast convergence of many observables with increasing model space. This
property is obviously very valuable when applying the method to studies
of many-body nuclear systems. The existence of multiple converging
sequences is another important property of the method; but one that has
not been extensively utilized in the past.
In this paper we have demonstrated the existence of multiple converging
sequences in the \emph{ab initio} NCSM, and we have discussed some
benefits of this property for certain applications. In particular, we
have proposed a revised strategy for computing cluster-approximated
effective interactions for $A$-body calculations in restricted model
spaces.

The fundamental principle that motivated this work is a particular
property of the theory of effective operators employed in the NCSM;
namely that convergence to the exact results is guaranteed for any
sequence that returns the bare Hamiltonian as the model space is
increased, $\nm \to \infty$. It was shown in Secs~\ref{sec:conve}
and~\ref{sec:resul} that the specific choice of employing the same
effective Hamiltonian for calculations in adjacent $A$-body model spaces
resulted in some attractive convergence properties. Firstly, it was
shown that it moderates the excursions in the renormalized results that
are usually encountered when performing calculations in smaller model
spaces, and secondly, that it gives a faster convergence of the relative
position of binding energies of opposite parity states. We also
demonstrated, by explicit large-scale calculations, that the revised
strategy indeed converges to the same result as the traditional choice
of using a $N_\mathrm{shift} = 0$ effective interaction for each model
space. We suggested that this new strategy is particularly useful for
studies of nuclear systems in which states of both parities are
simultaneously involved. Use of the revised strategy is specifically
envisioned for calculations of electromagnetic transitions between
states of opposite parities, as the theory of the corresponding
effective operator is simplified.

Another class of converging sequences was demonstrated by performing
calculations using the bare JISP16 realistic \nn\ interaction for a
series of HO frequencies. In this case, our approach obeys the
variational principle; for each choice of \ho\ we are guaranteed to
approach the exact result from above. In addition, all sequences are
guaranteed to converge to the same result as $\nm \to \infty$. We
utilized this property by carrying out fits to all sequences using
exponential functions, with the constraint to have a common
asymptote. This approach is potentially very useful for gaining
confidence in extrapolated results and for estimating their
uncertainties.

%
\begin{acknowledgments}
This work was partly performed under the auspices of the
U. S. Department of Energy by the University of California, Lawrence
Livermore National Laboratory under contract No. W-7405-Eng-48. One of
us (C.F.) acknowledges financial support from Stiftelsen Lars Hiertas
Minne and from Stiftelsen L\"angmanska Kulturfonden. One of us (J.V.)
acknowledges support from the U.S. Department of Energy Grant
No. DE-FG02-87ER40371. 
The calculations reported here were partly performed on computational
resources supplied by The Swedish National Infrastructure for Computing
(SNIC) under project SNIC007-07-56.
\end{acknowledgments}

\newpage
\printfigures
\end{document}